\documentclass[%
 reprint,
 superscriptaddress,
 prx,longbibliography,
]{revtex4-1}

\usepackage{graphicx}
\usepackage{dcolumn}
\usepackage{bm}
\usepackage{xcolor}
\usepackage[caption=false]{subfig}
\usepackage[tbtags]{amsmath}
\usepackage{hyperref}
\usepackage{pgf}
\usepackage{printlen}
\usepackage{braket}
\usepackage[normalem]{ulem}
\usepackage[colorinlistoftodos,prependcaption,textsize=tiny]{todonotes}

\newcommand{\red}[1]{{#1}}
\newcommand{\blue}[1]{{#1}}

\newcommand{\q}[1]{``#1''}

\makeatletter
\newcommand{\printfnsymbol}[1]{%
  \textsuperscript{\@fnsymbol{#1}}%
}
\makeatother

\begin{document}

\title{Gaussian Process States: A data-driven representation of quantum many-body physics}

\author{Aldo Glielmo}
\thanks{These two authors contributed equally to this work.}
\affiliation{Department of Physics, King's College London, Strand, London WC2R 2LS, United Kingdom}
\affiliation{Scuola Internazionale Superiore di Studi Avanzati (SISSA), Via Bonomea 265, 34136 Trieste, Italy}
\author{Yannic Rath}
\thanks{These two authors contributed equally to this work.}
\affiliation{Department of Physics, King's College London, Strand, London WC2R 2LS, United Kingdom}
\author{G\'{a}bor Cs\'{a}nyi}
\affiliation{Engineering Laboratory, University of Cambridge, Cambridge CB2 1PZ, United Kingdom}
\author{Alessandro De Vita}
\affiliation{Department of Physics, King's College London, Strand, London WC2R 2LS, United Kingdom}
\author{George~H.~Booth}
\email{george.booth@kcl.ac.uk}
\affiliation{Department of Physics, King's College London, Strand, London WC2R 2LS, United Kingdom}

\date{\today}

\begin{abstract}
We present a novel, non-parametric form for compactly representing entangled many-body quantum states, which we call a `Gaussian Process State'. In contrast to other approaches, we define this state explicitly in terms of a configurational data set, with the probability amplitudes statistically inferred from this data according to Bayesian statistics. In this way the non-local physical \red{correlated} features of the state can be analytically resummed, allowing for exponential complexity to underpin the ansatz, but efficiently represented in a small data set. The state is found to be highly compact, systematically improvable and efficient to sample, representing a large number of known variational states within its span. It is also proven to be a `universal approximator' for quantum states, able to capture any entangled many-body state with increasing data set size. We develop two numerical approaches which can learn this form directly: a fragmentation approach, and direct variational optimization, and apply these schemes to the Fermionic Hubbard model. We find competitive or superior descriptions of correlated quantum problems compared to existing state-of-the-art variational ansatzes, as well as other numerical methods.
\end{abstract}

\maketitle

\section{Introduction}
Representing entangled quantum many-body states efficiently and compactly is a major challenge, with
the need for tractable approaches underpinning a diverse set of fields including
quantum computation, electronic or nuclear structure, and quantum chemistry.
Advances in these fields are often punctuated by the discovery and exploitation of efficient wave
function forms for capturing certain physics with polynomial-scaling resources. These non-linear
parameterizations avoid the famed `exponential wall' in the complexity of the general solution.
This includes forms (and physical features) such as the Gutzwiller wave function \cite{Gutzwiller1963}
(suppression of local double occupancy), tensor network \cite{Schollwoeck2011,Orus2014,Sandvik2007} (low entanglement),
Laughlin \cite{Laughlin1983} (fractionalized excitations), or coupled-cluster
states \cite{RevModPhys.79.291} (low rank reducible correlations).
However, it is not just the quantum many-body problem which encounters a curse of dimensionality. Increasingly,
computational fields are turning to statistical inference and machine learning approaches to circumvent
this bottleneck, and to allow a more general framework to capture arbitrary functional forms
in high-dimensional spaces.
Following this spirit, in this work we espouse a new philosophy, whereby wave functions are not defined and
constrained by a fixed parameterization, but rather defined explicitly in terms of `{\em data}', with
their parameters considered statistically as random variables. This
leads to the description of entangled wave functions in a systematically improvable and non-parametric fashion.
The expressiveness, accuracy and compactness of the resulting state is found to be competitive or surpass
that of state of the art wave function parameterizations, encoding many important correlated states in a simple form.

The data used to define this wave function consists of a small `training' sample of underlying many-body
configurations and their (approximate) probability amplitudes. The question we wish to address first
is how to optimally use this training set to infer the wave function amplitudes on any other configuration
in the exponentially large Hilbert space, in a statistically rigorous fashion.
This allows us to define a complete wave function spanning all configurations, which can then be efficiently
sampled to extract any desired property of the system.
In this way, the state is explicitly parameterized by the data set, rather than this data used instead to optimize a fixed
functional form.
It is clear that this data-driven approach will not be possible for a random state where no structure exists
in the form of the amplitudes, but it is known that physical many-body states exhibit much structure
which can be effectively learnt in this approach, emerging from the underlying principles encoded in the Hamiltonian.
After this inferred wave function is defined and its limits discussed, we turn the idea into a practical and
general method for quantum problems, and demonstrate its accuracy and improvability in describing full $N$-body
quantum correlations by application to the Fermionic Hubbard model of the cuprates and strongly correlated solids.
In this work, we consider two sources for the data: configurations from the solution to a small fragment
of the full system, and a self-consistent approach to iterative selection and refinement of the data.
Furthermore, we show that an automatic selection of the relevant data required to capture the complexity of
the many-body correlations allows for a compact description of this state compared to other approaches,
with significant potential for a wide range of applications.

\section{Gaussian process regression for wave functions}
In order to describe the amplitudes of a quantum state, $\Psi({\bf x})$, evaluated for any
many-body configuration ${\bf x}$ in the underlying Hilbert space, we work within a
Bayesian inference framework known as Gaussian process (GP) regression. This approach falls under the umbrella of `kernel' methods, which constitute the alternate paradigm to neural networks in machine learning.
The great flexibility and ease-of-use of GPs and other kernel methods has brought them to prominence in a range of fields, including the interpolation of potential energy surfaces and other high-dimensional functions  \cite{Lilienfeld_12,Glielmo2017,Csanyi_10,Cheng_2019}. Advantages over neural networks include a rigorous underlying probablistic model and uncertainty bounds, the ease of incorporation of symmetries and conserved quantities, well understood regularization procedures, and robust and theoretically justified algorithms \cite{Rasmussen_book}.
In this non-parametric approach, an unknown target function is learnt from samples at a finite number of input points, allowing for a statistical inference of an estimator for the underlying function.
%
%
The weights of the model are considered as random variables, which define a distribution over all linear models of the state in a high-dimensional space of `features', defined by the map $\Phi({\bf x})$. The
scalar product of two configurations in this feature space define a kernel function, which can be evaluated between any two configurations, as $k({\bf x}, {\bf x'}) = \Phi({\bf x}) \cdot \Phi({\bf x'})$.
Estimating the expectation of the Gaussian process then involves evaluating a linear combination of kernel
functions \red{$k({\bf x}, {\bf x}_b')$} between the target configuration $\mathbf{x}$ and each configuration \red{$ \mathbf{x}_b' $} of the training database.

We introduce a `Gaussian Process State' (GPS) as the exponential of a GP estimator, where a configurational probability amplitude, $\Psi_g({\bf x})$, is given as
\red{
\begin{equation}
\Psi_g (\mathbf{x}) = \exp \left( \, \sum_{b=1}^{N_{b}} k(\mathbf{x}, \mathbf{x}_b') w_b \right),
	\label{eq:GP_wavefunction}
\end{equation}
}
\blue{where $w_b$ is the weight parameter of each data configuration defining the ansatz, which can in general be a complex number.}
%
%
Many existing parameterized states can be re-expressed in this form. For instance, the Gutzwiller Ansatz can be considered via a feature map into a single dimensional space, given by the number of doubly occupied sites in the configuration.
%
However, the key to circumvent the complexity of a general mapping is to note that the only quantity required for predictions of the configurational amplitudes is the scalar product of two feature vectors, defined by the kernel function.
%
%
By constructing the kernel function directly, the feature vectors are only ever implicitly defined, allowing for an arbitrary number of features to define the state, and a definition of the model explicitly and solely in terms of the
underlying data \red{configurations $\{ {\bf x}_b'\}$}.

This GPS is a `universal approximator', meaning that it can approximate a quantum state with arbitrary accuracy when provided with sufficient data.
A GPS can also be recast as a feed-forward neural network with a infinite-dimensional single hidden layer \cite{Neal94,lee2017deep}, connecting the approach to the recently developed neural-network states \cite{Jia2019,Gao2017,Carleo2018,Schuett2019,Zen2019,Choo2019,Hermann2019,Pfau2019,Clark_2018}.
These have recently come to prominence, particularly in the form of a `Restricted Boltzmann Machine' (RBM) which also forms a \red{non-parametric} and universal approximator for quantum states\red{, where `features' are effectively learnt variationally through the network architecture} \cite{Carleo2017,Melko2019,Nomura2017}.
The exponential form in Eq. \eqref{eq:GP_wavefunction} is reminiscent of other wave function forms, such as the Jastrow \cite{Jastrow} or coupled-cluster ansatzes \cite{RevModPhys.79.291}. However in contrast, the GPS is defined explicitly in terms of $N$-body configurations, rather than parameterized in terms of low-order features.
The exponential form also ensures that the overall state for our additive kernel is appropriately product separable in terms of its local features, and size extensive allowing its properties to appropriately scale to the thermodynamic limit.
As a scalar product of the configurational features, the kernel function \red{$k(\mathbf{x}, \mathbf{x}_b')$}, aims to quantify the similarity between any two configurations in the system, and it is clear that the specific choice of kernel function, as well as basis configurations, \red{$\{ {\bf x}_b' \}$} and weights \red{$w_b$}, is of paramount importance for the expressibility and success of the model.

\subsection{The kernel function}
The kernel function defines the set of \red{correlation} features for the state which are optimally modelled from the data, without explicitly applying the corresponding feature map, $\Phi({\bf x})$, into this space. 
With a kernel including all features we are able to reproduce any state with our ansatz, assuming we have a complete set of exact data points. 
More importantly though, we can introduce a set of kernels which gives rise to a systematically improvable ansatz which can accurately approximate the state using a far smaller set of training configurations. \red{This is achieved via the addition of hyperparameters, which weight the relative importance of fitting some features over others, but which still leaves an exact limit as the data set increases.}

We can write a configuration $\mathbf{x}$ in terms of its local Hilbert space on each spatial degree of freedom, $i$. As a specific example of Fermionic lattice models used later, we can consider four local Fock states for each lattice site, with $\mathbf{x} = x_i$ where $x_i \in \{\cdot, \uparrow, \downarrow, \uparrow \downarrow\}$ (for a spin or qubit model, there would just be two local states, $x_i \in \{ \uparrow, \downarrow \}$).
The kernel function compares these local occupations between two configurations, extracting multi-site \red{correlation} features common to both.
To do this, the kernel $k(\mathbf{x}, \mathbf{x}')$ is written as a linear combination of products of delta functions $\delta_{x_i x_j'}$ which we define to be $1$ if the local occupation $x_i$ is equal to the local occupation $x_j'$, and 0 otherwise.
We do not assume that the dimension of the two configurations are the same, and can therefore compare configurations with different numbers of lattice sites.

The simplest kernel just compares the local occupancies, given by
\red{
\begin{equation}
  k^{(1)}(\mathbf{x}, \mathbf{x}') = \sum_{i}^L \delta_{x_i x_j'},
  \label{eq:k1_kernel}
\end{equation}
where $L$ refers to the size of the lattice corresponding to $\mathbf{x}$ (denoting the test amplitude to infer) and $x_j'$ denotes the local occupancy of the training point at a chosen reference site $j$. The reference site for the training lattice can be chosen arbitrarily for translationally symmetric systems, since there always exists a potential training configuration which can describe all local environments of each site.
Additional summation over the sites of the training lattice in Eq.~\ref{eq:k1_kernel} is possible, to ensure translational symmetry in the training data. However, this was not found to improve the results sufficiently given the additional cost to evaluate, so only the sum over the test lattice is included, ensuring the wave function still retains overall translational symmetry.}
The superscript `$(1)$' emphasises that this kernel extracts features corresponding to the Fock space of local, single-site plaquettes, therefore implicitly mapping each configuration to a four-dimensional space corresponding to its number of each local state.
This therefore extracts the features used in the Gutzwiller ansatz (number of doubly occupied sites), with the only difference that the \red{parameter} of the feature is given implicitly by the data amplitudes, rather than explicitly in this feature space as is traditionally done.

\begin{figure}[h]
  \centering
  \includegraphics[width=\columnwidth]{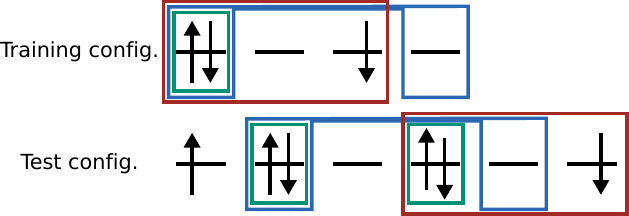}
  \caption{Exemplification of the kernel function feature space. The kernel counts \red{correlation} features common to the test configuration (bottom) and training configuration (top), including e.g.\ the single-site Gutzwiller feature (green), two-site features explicit in the Jastrow wave function (blue) and nearest neighbour plaquette features used in EPS ansatzes (red).} \label{fig:kernel_visualisation}
\end{figure}

In order to include higher order \red{correlated} features, we can include comparisons of the occupations of plaquettes over multiple sites for the two configurations (see figure~\ref{fig:kernel_visualisation}).
Here, the advantage of using a kernel function to extract descriptors implicitly becomes very prominent; while the number of possible multi-site features grows exponentially with the number of sites in the plaquette considered, the kernel can still be efficiently evaluated. Multi-site features can simply be included via products of delta functions evaluated for different displacements of sites $i$ and $j$.
Exploiting this, we can analytically resum the contribution to the kernel function from all possible topologies and ranks of \red{correlated} descriptors defined up to a maximum number of sites (controlled by the hyperparameter $p$), and maximum length (controlled by a set of displacements $D$), giving a general kernel form
\red{
\begin{align}
  k(\mathbf{x}, \mathbf{x}') &= \sum_{i}^L  \delta_{x_i x'_j} \, \left( \frac{\tilde{k}(\mathbf{x}_i, \mathbf{x}'_j)}{\sqrt{\tilde{k}(\mathbf{x}_i, \mathbf{x}_i) \tilde{k}(\mathbf{x}'_j, \mathbf{x}'_j)}} \right)^p \label{eq:kernel}\\
	\tilde{k}(\mathbf{x}_i, \mathbf{x}'_j) &=  \theta + \frac{1}{p} \sum_{d \in D}  \delta_{x_{d(i)}  x'_{d(j)}} f(d).
  \label{eq:kernel_local}
\end{align}
}
%
\red{In this kernel, the local occupancy over all sites $i$ of the test configuration is compared to the occupancy of reference site $j$ in the training configuration; if the local occupancy is the same ($\delta_{x_i x'_j}=1$), then the local surroundings (of $\mathbf{x}_i$ and $\mathbf{x}'_j$) around sites $i$ and $j$ are also compared by means of the `local' kernel $\tilde{k}(\mathbf{x}_i, \mathbf{x}'_j)$.}
%
The local comparison includes all displacements $d \in D$ of $i$ and $j$, running over the $\vert D \vert$ closest sites.
\red{In this work, we choose $D$ to include all possible displacements over the lattice the training configurations $\mathbf{x}'$ are associated with.}
%
Raising the local kernel to the power of $p$, we obtain a linear combination of all possible multi-site \red{correlation} features in our implicit parameterization of the state, in the form of an `additive' kernel, which hierarchically includes all lower-rank features \cite{AdditiveGP}. As $p \rightarrow (L-1)$, the implicit feature map of each configuration, $\Phi({\bf x})$, is injective, fully characterizing the configuration.
%
\red{Further flexibility in this form is provided by the hyperparameter $\theta$, controlling the relative importance of fitting high-order vs low-order features, as well as the function $f(d)$, which in this work, unless otherwise stated, we choose to be the inverse of the displacement distance, $f(d)=1/|d|$, to weight the importance of fitting short-range correlation features more than longer-range. This distance weighting will be further generalized to add more flexibility in Sec.~\ref{subsec:bootstrapping}.}
\red{The complexity of the underlying parameterization of the GPS is therefore controlled by hyperparameters $\theta$ and $p$.}
These will optimally model the \red{correlated} features in the data, describing entanglement ranging up to $p+1$ sites, with $\theta$ controlling the relative importance in the rank of these features.

The introduced kernel can be shown to model well known ansatzes under specific limits of these hyperparameters.
For $\theta \to \infty$ or $p \to 0$, we recover the $k^{(1)}$ kernel introduced in Eq.~\eqref{eq:k1_kernel} which gives rise to a purely local Gutzwiller underlying form. For $\theta = 0$ and $p=1$, we recover a generalised two-body Jastrow factor, and finally $\theta = 0$ and $p \rightarrow \infty $ generates an Entangled Plaquette State (EPS) \cite{Changlani2009,Mezzacapo_2009} with all plaquette sizes up to $\vert D \vert +1$.
As $D$ is then allowed to increase in size to include all system displacements, then the kernel is at its most expressive, allowing all features to be considered, and an underlying (over-)parameterization of the complete wave function space.

Other important wave functions are also compactly expressed within a GPS formalism. The $n$-qubit W state \cite{Duer2000} is a multipartite entangled state, important in quantum computing, which cannot be efficiently represented as an EPS \cite{MPSRepr_07}.
This can however be represented as a GPS ($\theta = 0$, $p \rightarrow \infty $) with a single data configuration.
Similarly, strongly-interacting states such as the Laughlin wave function \cite{Laughlin1983, Glasser2016} describing the topologically-ordered fractional quantum Hall effect can be exactly represented as
a product of two GPS forms acting on a single data configuration.
A full derivation of these limiting cases of the kernel are found in Appendix~\ref{subsec:SM_kernel_limits} and ~\ref{subsec:Representatbility}.

Symmetries of the system can also be simply included in the kernel form. For instance, to incorporate the spin symmetry of the Hubbard model considered later into our ansatz, we can use the symmetrized kernel
\red{
\begin{equation}
  \bar{k}(\mathbf{x}, \mathbf{x}') = k(\mathbf{x}, \mathbf{x}') + k(\mathbf{x}, \bar{\mathbf{x}}'), \label{eq:spinsym}
\end{equation}
}
where $\bar{\mathbf{x}}'$ represents the spin-reversal of configuration $\mathbf{x}'$.
Key to the performance of the GPS, is that despite the fact that an exponential number of \red{correlation} features can underpin the state, the kernel presented in Eq.~\eqref{eq:kernel} can be computed efficiently in only
\red{$\mathcal{O}[L \, \vert D \vert]$} time (which can be reduced to
\red{$\mathcal{O}[L]$}
when only a local change to ${\bf x}$ is made to a known kernel value, as is common when sampling the space).
A full configurational amplitude evaluation will require \red{$N_b$} kernel computations, as given in Eq.~\eqref{eq:GP_wavefunction}.
This construction of the state in the space of $N-$body data configurations therefore allows for its flexibility and expressibility.
We now turn to the consideration of how compact the set of data configurations can be, and \red{Bayesian framework for} computation of their selection and optimization of weights as key for the efficiency of the model.

\subsection{Training and selection of data}
Similar to the constructive algorithm for a matrix product state via successive SVD and decimation steps, we aim to devise a deterministic algorithm which allows for the compression of an arbitrary state into GPS form. This requires selection of a set of training configurations, and associated trained weights to optimally represent the original wave function.
Ideally, we want this training set to be \emph{sparse} for a compact description of the state, by exploiting the underlying structure of the \red{dominant correlation} features of the target wave function.
The required size of the training set will depend on the choice of kernel hyperparameters and desired accuracy in reproducing the state, \red{as given by the variance of the likelihood estimator}.
In particular, the quality of the wave function can be improved if a more complex kernel is used with a larger set of \red{correlation} features (e.g.\ by increasing $p$).
However, we expect this will consequently require a larger training set of configurations to appropriately fit - a common tenet of machine learning.

We base our approach on the \emph{Relevance Vector Machine} (RVM) \cite{Tipping2000}, a Bayesian algorithm for sparse learning.
%
Within the RVM, the weight of each of a large set of candidate training configurations is considered as separate random variables, with Gaussian prior distribution.
%
The variances of such prior distributions can be efficiently and deterministically optimized through the maximization of the marginal likelihood of the model \cite{Tipping2003}. \red{A large optimal variance for the configuration indicates its relative importance in reproducing the wave function amplitudes to a desired fidelity (given by $\tilde{\sigma}^2$ in App.~\ref{subsec:RVM}), and is therefore selected in the data set.}
%
%
%
Concurrently, the optimal weights of the resulting configurations, \red{$w_b$}, are given by the mean of the posterior distribution for each data point.
Complex Gaussian forms for this distribution allow for the optimization of complex weights, which admits signed amplitudes in the overall model.
%
%
This yields a \red{rigorous Bayesian} approach to automatically pick a highly compact training set of support configurations and optimized weights for the GPS, in order to faithfully reproduce a state to a desired accuracy. This is a systematically improvable approach to GPS construction, whereby increasing the complexity of the kernel and hence underlying features parameterizing the state, \red{or reducing the variance of the likelihood, $\tilde{\sigma}^2$}, automatically results in a corresponding increase in the data set size selected by the RVM. Further details on this statistical approach for selecting and optimizing the data of the GPS is given in Appendix~\ref{subsec:RVM}, \red{while a more in-depth exposition of this Bayesian approach can be found in Ref.~\onlinecite{rath2020bayesian}}.

\begin{figure*}[t!]
  \centering
  \includegraphics[width=0.9\textwidth]{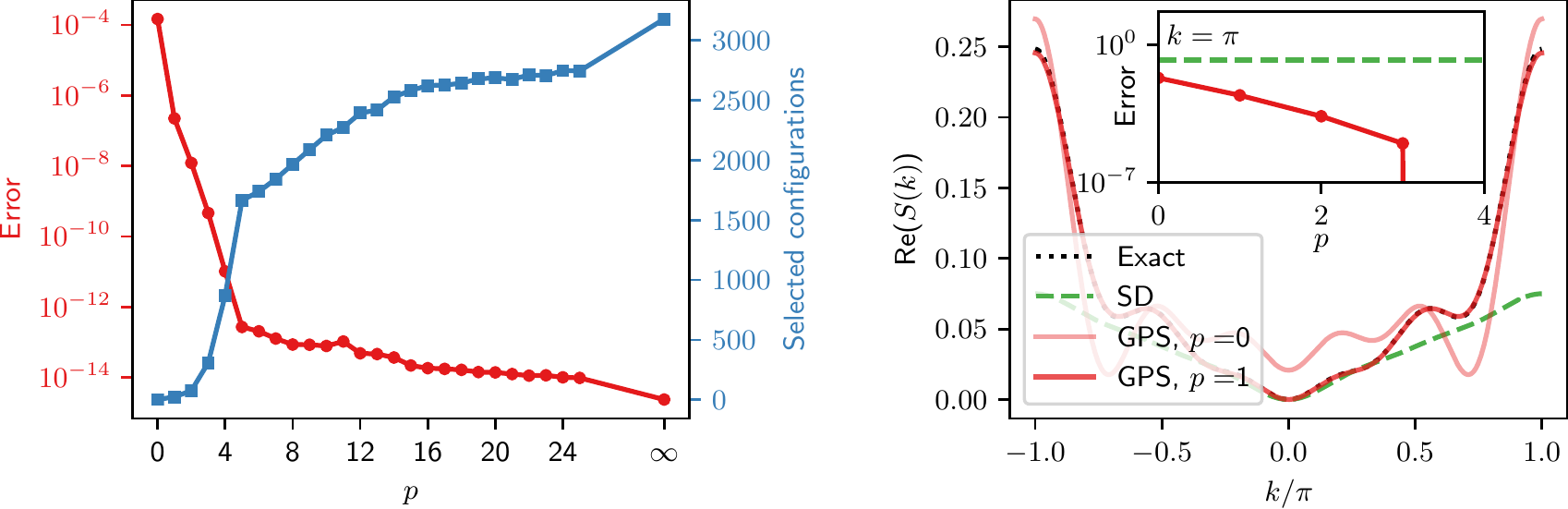}
  \caption{\red{Left panel:} Mean squared GPS amplitude error for the ground state of the 10-site, half-filled 1D Hubbard chain at $U=8t$, as the kernel power hyperparameter is varied ($p$, with $\theta = 0$). Also shown is the size of the training set of configurations selected by the RVM from the exact data to construct the GPS \red{with this underlying kernel complexity}. \red{Right panel: Real part of the spin structure factor $\textrm{Re}(S(k))$ corresponding to the exact wavefunction (black, dotted), the GPS with $p=0$ and $p=1$ (red, solid) and the single Slater determinant representation obtained with the Hartree-Fock method (green, dashed) for the same system. The inset shows the absolute error of $\textrm{Re}(S(k=\pi))$ for the GPS with respect to the kernel hyperparameter $p$ as well as the error obtained for a single Slater determinant representation.}}
  \label{fig:interpolation_1D}
\end{figure*}

We demonstrate the compression of a correlated state to GPS form by considering the paradigmatic Fermionic Hubbard model, described by the Hamiltonian
\begin{equation}
	\hat{H} = -t \sum_{\langle i,j\rangle,\sigma} {\hat c}_{i,\sigma}^{\dagger} {\hat c}_{j,\sigma} + U\sum_{i} \hat{n}_{i \uparrow} \hat{n}_{i \downarrow}.
\end{equation}
\red{The left panel of Fig.}~\ref{fig:interpolation_1D} shows the fidelity in compressing the ground state of this model to GPS form as the kernel complexity is increased via the $p$ hyperparameter, for a 10-site system which can be exactly solved.
The error is defined as the mean squared error over all configurations, while the corresponding size of the automatically selected data set is also shown.
With $p=0$, only single-site local \red{correlation} features are represented, resulting in the selection of only
two relevant configurations to specify the state with this level of description, as the entire (four-dimensional) feature space is rapidly saturated by the data.
As the kernel complexity increases, larger training set sizes are chosen, with a rapid corresponding decrease in the error of the inferred GPS.
%
%
\red{For the most complex kernel containing all range and rank of correlations, up to $\sim 3200$ configurations are selected in the training set, resulting in the exact wave function reproduced as a GPS to numerical precision.}
This represents a key feature of the GPS, that a desired kernel complexity can be used to automatically construct a sparse training set and weights required to optimally represent the complexity of correlations in the state.
It is encouraging that the convergence of the GPS is so rapid with training set size, only requiring \red{$76$} training configurations to represent all amplitudes with a mean squared error of $\sim10^{-8}$ ($p=2$) for this system.
%

\red{As the GPS directly models the wave function, all physical properties of the state can be extracted.
This is exemplified in the right panel of Fig.~\ref{fig:interpolation_1D} which shows the spin structure factor $S(k) = \frac{1} {L} \sum_j e^{-i k r_j} \langle \hat{S}_0 \cdot \hat{S}_j \rangle$ 
obtained for the GPS with $p=0$ and $p=1$ in comparison to the exact state, and the mean-field description.
Although it only comprises $24$ data configurations, the GPS with $p=1$ already yields an accurate description of the structure factor which is significantly better than that of a single Slater determinant.
The inset shows the absolute error of the structure factor at the boundary of the first Brillouin zone as a function of $p$.
Similar to the mean squared error on the exact wave function amplitudes, the structure factor error systematically decreases with increasing values of $p$.
Already the GPS with $p = 4$ matches the exact result up to numerical precision which indicates the efficient description of even non-local correlations in this description.}
\begin{figure}[h]
  \centering
  \includegraphics[width=\columnwidth]{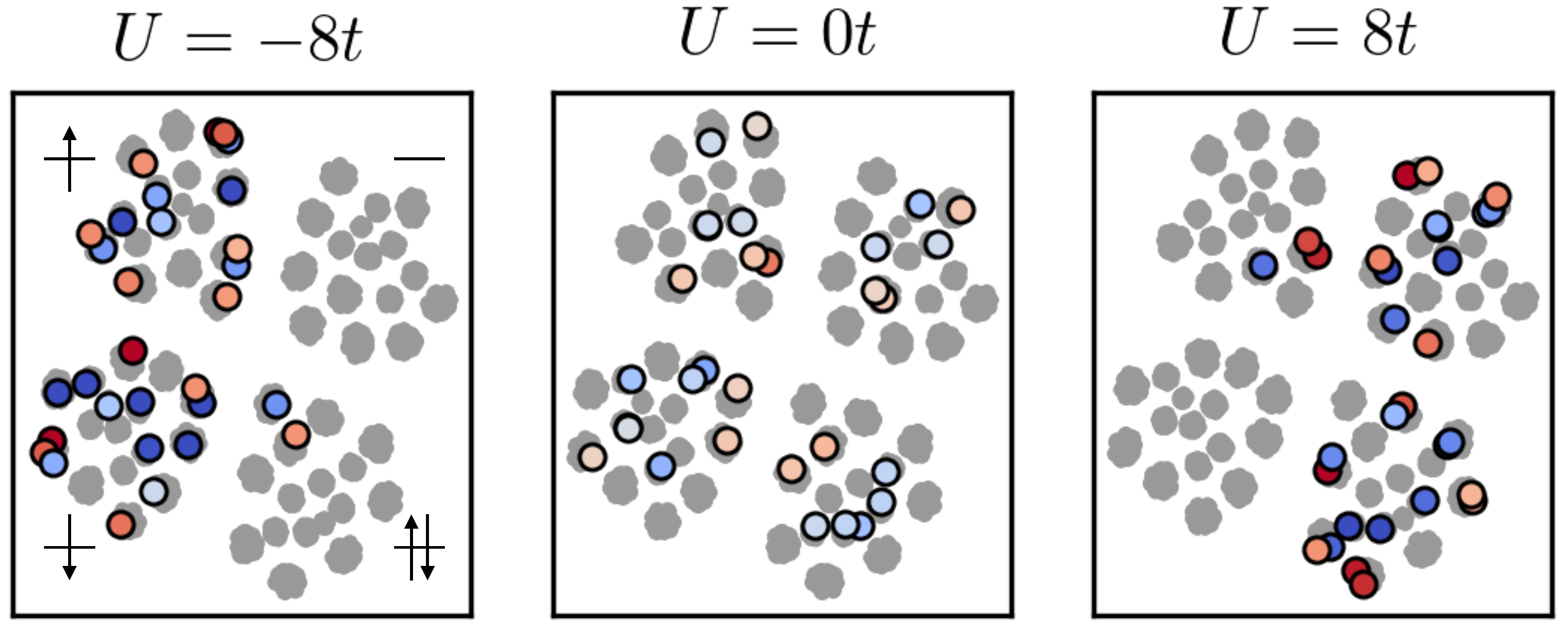}
  \caption{
  \red{
  Two dimensional t-SNE projection of the Hilbert space of a half-filled 1D Hubbard chain of 8 sites.
  The four clusters are specified by the occupancy of the reference site of the configurations (clockwise from the top left: spin-up electron, holon, doublon, spin-down electron). The coloured points are the 30 configurations selected by the RVM as the most relevant, with red and blue colors indicating positive and negative weights.}
  }
  \label{fig:rvm_visualisation1D}
\end{figure}

\red{The GPS model can also provide a unique insight into the underlying emergent physics of a state, through an analysis of the compact set of relevant configurations selected, and their corresponding dominant features.
A dimensionality reduction \blue{t-distributed stochastic neighbour embedding (t-SNE)} algorithm can be used to project the data configurations onto a two-dimensional plane, while optimally preserving a distance associated with the kernel function \cite{Maaten2008}.
%
%
Figure~\ref{fig:rvm_visualisation1D} shows this for all configurations under the $p\rightarrow \infty$ kernel given in Eq.~\ref{eq:p_infty_kern}. The full configurational space is shown as individual grey points, which naturally cluster together into four main groups based on the local Fock occupations of the reference site (up-spin, down-spin, unoccupied and doubly occupied). Self-similar clusterings then also emerge in a clear hierarchical manner within each of these main clusters, based on their increasingly long-range correlation features in common. However, the selected configurations chosen as basis configurations to support the GPS heavily depend on the dominant physics and correlations emerging from the model.}

\red{The optimal 30 configurations automatically selected by the RVM are colored from dark red (large positive weights) to dark blue (large negative weights). 
For negative $U/t$ (the attractive Hubbard model), the RVM mostly selects configurations with single occupancies on the reference site, suppressing their amplitudes relative to the uniform state, resulting in an enhancement of short-range doublon formation and emergent pairing order (the symmetry of Eq.~\ref{eq:spinsym} ensures that the state is invariant to global spin flips of the data). Conversely, at large positive $U/t$, doublon/hole occupancy is predominantly suppressed, resulting in the emergence of anti-ferromagnetic features from the chosen data points. At $U=0$, data is selected uniformly through the Hilbert space in order to build up the long-range delocalization of the quasiparticles in the metallic state. Further insight into the subtle nature of the correlated physics emerging from these states can be found by consideration of the longer-range features within each chosen data configuration, giving a new perspective into emergent correlated physics within the GPS framework.}

\red{If the relevant physics in the wave function does not substantially change with lattice size, then we can reliably assert that the error per particle will remain constant as the system grows for the same fixed configurational data set and kernel complexity. While this will only hold rigorously for systems beyond their bulk correlation length, we now use this observation within a practical construction of a GPS in the large system limit, by inferring from the truncated features of a smaller lattice.}

%

\section{Practical Gaussian Process States}
\subsection{Extrapolated GPS}
While the compact representation of a GPS has been shown to have a number of promising features, we turn to practical algorithms for quantum many-body problems, where the desired state is not generally known in advance. This requires a method to choose a necessarily approximate initial state to compress to GPS form, before potential iterative refinement. We first consider an approach whereby we can exactly solve only a fragment of the system, and use the information encoded about the correlations within that fragment to infer the wave function amplitudes in GPS form on a larger simulated system - an approach which has recently also been considered for neural network states \cite{Bagrov19}. 
We denote this fragment the `training' lattice, and the full system of interest the `simulation' lattice (e.g. in the thermodynamic limit).
The approach is possible as the kernel function can compare \red{correlation} features on training and simulation configurations in different sized Hilbert spaces, as well as the multiplicatively separable features of the GPS ensuring a size extensive and translationally invariant overall description.
It is also important to ensure that single-particle physics beyond the lengthscale of the training lattice is still incorporated, which is done via a simple, fixed free particle product state, which has the exact sign structure for this model. The full wave function on the simulation lattice is therefore the product of this free-fermion state and a GPS defined with a training data set on a small sub-lattice, which is designed just to infer the short-ranged \red{correlated} corrections to the underlying single-particle description. This mirrors the use of other successful fragmentation approaches for strong correlation, such as dynamical mean-field theory, where correlations in a small fragment are replicated across a larger system \cite{Georges96}. However here, no symmetries are broken in the state, and a correlated wave function is defined over all configurations.

\begin{figure}[t]
  \centering
  \includegraphics[width=\columnwidth]{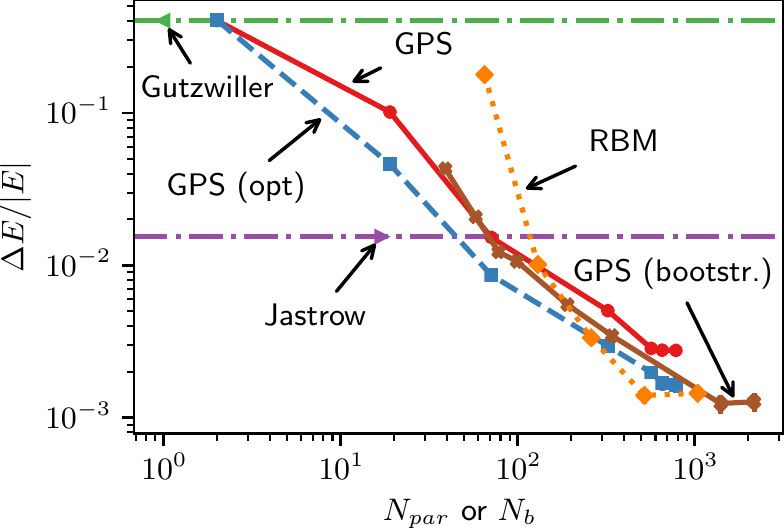}
  \caption{\red{Relative energy errors for the half-filled Hubbard chain \blue{(compared to DMRG)} at $U=8t$, $L=32$, as a function of the size of the chosen set of configurations (\red{$N_b$}), as the kernel complexity $p$ is increased from $0$ to $6$ ($\theta=0, L'=6$). For other methods, $N_{par}$ denotes the number of variational parameters optimized in the state, as a measure of compactness and variational freedom of the state. The GPS is trained on the features of the ground state of a small $12$-site lattice using configurations with all particle and spin fluctuations from a $L'=6$ lattice. Subsequent variational optimization of the training weights of the GPS gives rise to the `GPS (opt)' values. For comparison, variationally optimized and translationally-symmetric Gutzwiller, Jastrow and RBM wave function results (with the same fixed free-Fermion state) are included, with the RBM ratio of hidden to visible nodes ($\alpha$) chosen as $1$, $2$, $4$, $8$ and $16$ \cite{Carleo2017}, with the best result of $10$ independent seeds taken.
  Also included are results from iterative bootstrapping of the GPS, as introduced in Sec.~\ref{subsec:bootstrapping}.
  The bootstrapped results were obtained by starting from a random initial set of data configurations.
  For these data points, the $p \to \infty$ limit of the kernel was used where the scaled distance weights $f(d)/\theta$ in the kernel function were optimized variationally.}}
  \label{fig:extrapolation_1D_U2_U8}
\end{figure}
We use this approach to model the ground state of the Hubbard model for larger lattices, where the energy per site is well converged. In 1D, numerically exact results are accessible via the DMRG \cite{Chan12,Schollwoeck2011}, allowing rigorous benchmarking of the accuracy of the resulting GPS state.
\red{For this work, we choose an $L'= 6$ site training lattice for the configurations $\mathbf{x}'$ and learn the model by training on the exact wave function data of a $12$-site system which can be solved exactly.
By considering training configurations with all possible particle number and spin fluctuations, it is ensured that a complete set of all features in the six-site range is spanned when fitting the $12$-site data with the other sites outside this range effectively acting as an `entanglement bath' to ensure a representation in the data of all the six-site features.} 
%
\red{There is no fundamental difficulty in the use of basis configurations to support the GPS defined on a different lattice than the one we used to define the training amplitudes, due to the sparse GPS approach taken to the fitting as detailed in App.~\ref{subsec:RVM}.
We then use the Bayesian fitting via the RVM to select the training set for different values of $p$ by fitting the exact solution, without any further optimization or adjustable parameters. This results in an exceptionally computationally inexpensive approach.}
%
Since the resulting GPS on the simulation lattice \red{($L = 32$ sites in this example)} is easily specified, a simple Markov chain Monte Carlo is performed to sample the energy of the state, without a sign problem, 
giving the results \red{of the red line} in Fig.~\ref{fig:extrapolation_1D_U2_U8}.

For comparison, using standard variational Monte Carlo methodology developed in the `mVMC' package \cite{PhysRevB.64.024512,Misawa2019}, we also optimize translationally-symmetric, real-valued Gutzwiller, Jastrow and RBM states on top of the same (fixed) free-Fermion state, where the variationally lowest result over multiple independent seeds is shown, to allow for faithful comparison of the accuracy and compactness of these states.
For the GPS, $p=0$ is shown to be essentially Gutzwiller accuracy, while $p=1$ is slightly worse than the Jastrow description due to the neglect of two-site correlated features beyond a distance of six sites.
Increasing $p$ increases the quality of results up to $p=4$, at which point
\red{all correlations required to describe} the lengthscales of the training lattice are captured, resulting in a relative energy error of 
\red{$\sim3\times10^{-3}$}, significantly beyond the accuracy of traditional Slater-Jastrow approaches, with a chosen data set of \red{$\sim570$} configurations.
To mitigate the fact that the parameters of the model (the weights of the data) were deterministically trained by the RVM on the solution of a small sub-lattice, we can also variationally optimize these weights for the overall state keeping the chosen data configurations fixed (blue squares in Fig.~\ref{fig:extrapolation_1D_U2_U8}).
This removes the bias from the weights having been optimized on the smaller lattice, but the improvement is rather marginal. The ultimately accuracy is comparable to the best RBM results, where increasing the number of hidden nodes allows for an in principle exact description of the state.
The remaining error therefore primarily stems from the lack of \red{correlated} features in the kernel with a range beyond six sites, for which a larger training lattice for the data is required, and will be addressed in the next section.

\begin{figure}
  \centering
  \includegraphics[width=\columnwidth]{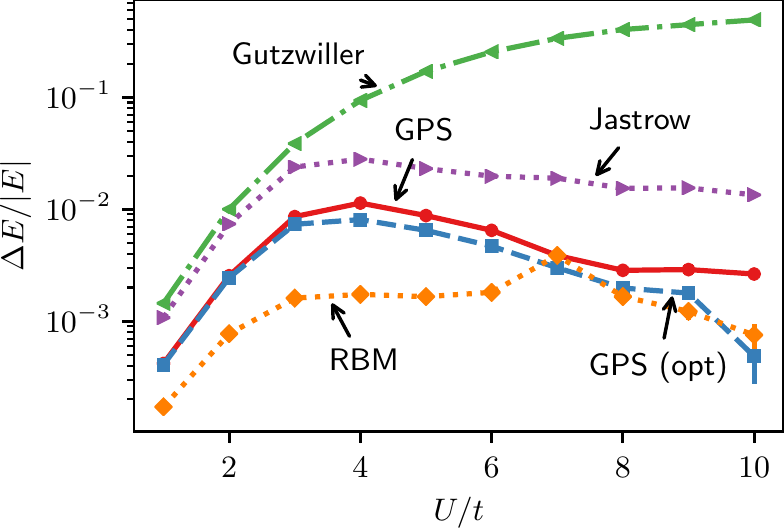}
  \caption{
  \red{Relative energy error (compared to DMRG) against $U/t$ for GPS, Gutzwiller, Jastrow and RBM states, for a half-filled 32-site Hubbard chain. The GPS ansatz is obtained from a fit on a $12$-site chain solution with training configurations coming from a $L' = 6$ lattice with all possible particle and spin fluctuations using kernel hyperparameters ($p=4$, $\theta=0$).}
  All ansatze are built on the same fixed free-Fermion orbitals. Monte Carlo optimizations for the Gutzwiller, Jastrow and RBM \red{(with $\alpha =5$)} states are performed for 
  $10$ seeds, with the variationally lowest result shown. The red curve shows the GPS result obtained with the weight solely obtained from the fit on the fragment amplitudes, with the blue curve representing additional variational optimization of the weights, with the $L'=6$ data set fixed.}
  \label{fig:energy_vs_U_1D}
\end{figure}
Compact `extrapolated' GPS descriptions with competitive accuracy compared to other variational ansatzes can be found across a range of correlation strengths, as demonstrated in Fig.~\ref{fig:energy_vs_U_1D}.
%
\red{We take the $p=4$ GPS, with the same setup as Fig.~\ref{fig:extrapolation_1D_U2_U8}, and consider the accuracy of the simple extrapolation procedure as the effective Coulomb strength, $U/t$, is varied.}
%
While the character of the ground-state changes significantly in this range, the GPS results are consistent, outperforming Gutzwiller and Jastrow parameterizations for all $U/t$, even without any variational optimization of the weights.
\red{The $\alpha=5$ RBM results are
superior for $U \leq 6t$, where the correlation lengths exceed the size of the training lattice features, however the increasing dominance of short-range correlations at larger $U/t$
gives GPS results with similar accuracy to the shown RBM results.
%
%
However, it is clear that to return to a systematically improvable description, it is necessary to go beyond the constraints imposed by a small training lattice, and to include correlated features at all lengthscales and rank in an iterative approach.}

\subsection{Bootstrapped Optimization of GPS}
\label{subsec:bootstrapping}
\begin{figure}
  \includegraphics[width=\columnwidth]{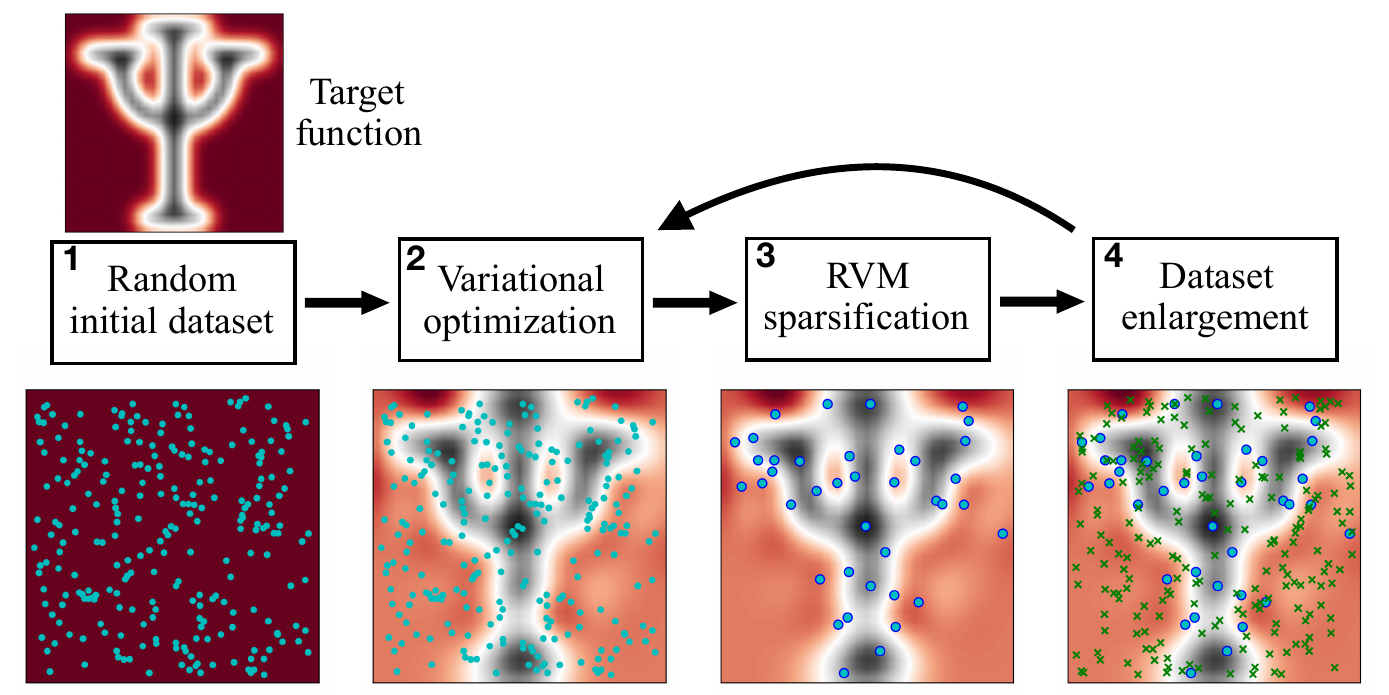}
  \caption{Illustration of the bootstrapping algorithm for the 2D target function shown in the top left corner. 1) An initial set of configurations is randomly selected (blue points), with $w_t=0$ ensuring no features are described. 2) The weights of each configuration are optimized variationally, reasonably describing the desired function. 3) The RVM is used to drastically sparsify the dataset without changing the learned function. 4) Configurations with high local energy variance are added to the sparsified set (green crosses). Note that the final three panels model the function to essentially the same accuracy, allowing systematic improvement in the refining of the data set.}
  \label{bootstrap_illustration}
\end{figure}

We propose a further numerical procedure for a series of optimization, compression and data reselection steps to iteratively refine a GPS and its data set, illustrated for modelling a continuous function in Fig.~\ref{bootstrap_illustration}. The approach is defined directly on the target system, and therefore avoids the need to define a fragment system or restrict to short range features.
%
\red{If we do not have access to any previous wave function data, we initialize the GPS ansatz with a random set of data configurations with \red{$w_b=0$}.
Alternatively, we may simply pre-train the GPS by selecting the initial data configurations and weights from compression of a previously known approximate wave function, e.g. an optimized Jastrow ansatz, into GPS form. This step can give an improved starting point for the subsequent variational optimization of the ansatz.}
\red{After the GPS ansatz is variationally optimized,} the data set required for the optimized GPS is sparsified via the RVM procedure, which prunes redundant data from the original specification which are not required to describe the state to the target accuracy (as given by the kernel complexity \red{and the noise parameter $\tilde{\sigma}^2$ regularizing the fit of the data}).
This optimized GPS is then augmented with additional data points outside the original set, in order to acquire additional flexibility, before all weights are then reoptimized variationally and the process is repeated until convergence.
\red{In the last step, we do not add more data configurations after the RVM has pruned non-relevant configurations from the model but just run an additional variational optimization to obtain the final representation.}

There are a number of possible criteria for selection and addition of new trial data configurations to improve the GPS; the largest uncertainty in the GPS (obtained in a Bayesian framework), highest local energy, or largest contribution to the variance of the local energy.
All these criteria indicate that the configuration is relatively poorly represented in the current GPS data set, which would benefit from explicit inclusion and optimization of these points. We choose the latter criterion, with the selection made from those configurations sampled during the previous Monte Carlo optimization of the GPS.
\red{We choose the number of configurations added to the data set to be a constant fraction (in this work 25\%) of the size of the compressed data set.}
\red{This ensures that a large fraction of the data at any point has been already deemed relevant by the RVM, whilst allowing continued exploration of the relevant data and feature space.
The complexity of the model (and therefore its accuracy) can be influenced by tuning the noise parameter $\tilde{\sigma}^2$ used to regularize the RVM fit of the data (see Appendix~\ref{subsec:RVM}).
Choosing a smaller value of $\tilde{\sigma}^2$ enforces a higher fidelity in the reproduction of the wave function at each sparsification step, with a commensurate increase in required data set size.} 

\red{The performance of the bootstrapped algorithm (without any pretraining) is compared to the extrapolated GPS method as brown crosses in Fig.~\ref{fig:extrapolation_1D_U2_U8}, for the $L=32$ site system at $U=8t$.
To avoid restriction of the feature space, we use the $p \to \infty$ limit of the kernel (see Appendix~\ref{subsec:SM_kernel_limits}) which takes the form
\begin{equation}
    k_{p \to \infty}(\mathbf{x}, \mathbf{x}') = \sum_{i}^L \delta_{x_i x_j'} \, e^{-\sum_{d \in D}  (1-\delta_{x_{d(i)} x_{d(j)}'}) \frac{f( d )}{\theta}} \\
	 \label{eq:p_infty_kern}
\end{equation}
and additionally variationally optimize the scaled distance weighting in the kernel of $f(d)/\theta$ to provide more flexibility in the distance weighting of the feature space.
The total number of variational parameters of the ansatz reported in the figure is therefore $N_b + L - 1$, where $N_b$ is the final number of data configurations reached in the bootstrapping, and $L$ is the lattice size. We note that the complete specification of a GPS requires more than just these parameters, since the details of the discrete basis configurations to support the state are also required. However, the RVM procedure to select these configurations is highly efficient, such that the picking of these configurations is a negligible part of the simulation protocol.
The bootstrapping procedure, along with the optimization of the distance weighting of the features in the kernel allows us to reach representations which systematically surpass the accuracy obtained from the fragmentation procedure used previously, as all long-range features are included in the description. This allows us to reach a relative energy error of 
$\sim 1.2 \times 10^{-3}$ using $N_b=1369$ data configurations.
}

\begin{figure}
  \centering
  \includegraphics[width=\columnwidth]{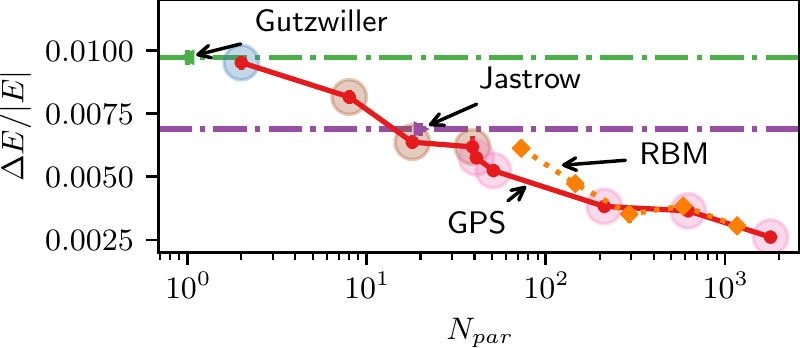}
  \caption{Energy errors obtained for the bootstrapped GPS for the half-filled $6 \times 6$ 2D square Hubbard lattice at $U=8t$ \blue{(compared to benchmark AFQMC values \cite{Qin2016})}, plotted against number of variational parameters (excluding those in the Pfaffian state). \red{Results are systematically improved via use of a more complex kernel, or decrease of the ${\tilde{\sigma}}^2$ regularization parameter, which increases the chosen data set size. The GPS with blue background shading was obtained with the $k^{(1)}$ kernel, values with brown shading correspond to the $p \to \infty$ kernel where the distance weighting was fixed at $f(d) = 1/ \vert d \vert$ and $\theta$ was variationally optimized, while for the pink highlighted results, all scaled distance hyperparameters $f(d)/\theta$ were optimized variationally (giving $N_b + 35$ variational parameters). Also displayed for comparison are the energy errors obtained with the Gutzwiller, Jastrow and RBM ansatzes (with $\alpha$
  chosen as $1$, $2$, $4$, $8$, $16$). Each of these displayed comparison values corresponds to the lowest energy obtained for repeated calculations with five different random seeds.}}
  \label{fig:bootstrapping}
\end{figure}

We also apply this `bootstrapped' GPS approach to the 2D square Hubbard lattice at $U=8t$, where the electronic structure problem is far more challenging.
\red{We consider the quality of the bootstrapped GPS for three different forms of the kernel: the restrictive $k^{(1)}$ kernel of Eq.~\ref{eq:k1_kernel} (obtained as $\theta \to \infty$), the $p \to \infty$ kernel of Eq.~\ref{eq:p_infty_kern} where the hyperparameter controlling the relative importance of fitting high vs. low-rank features ($\theta$) is variationally optimized but where the distance weighting is fixed at $f(d)=1/|d|$, and finally the kernel where this distance weighting is also entirely optimized. Further fine control of the accuracy vs. compactness of the GPS can be achieved via the regularization parameter ${\tilde{\sigma}}^2$ within each of these kernel choices, defining the fidelity in reproducing the state upon the compression step, and giving a clear route to improvability in the GPS results \cite{rath2020bayesian}.}
%
%
We improve the flexibility of the orbital part of the wave function by using a simultaneously optimized Pfaffian state, together with translational and spin-symmetry projections on this part of the wave function \cite{Misawa2019, Tahara2008, Mizusaki2004}.
This allows for more freedom to alter the nodal structure, and we restrict the weights of the GPS and RBM forms to real values.

Fig.~\ref{fig:bootstrapping} shows the relative energy error of the bootstrapped GPS with respect to the number of variational parameters used, as the data set size increases and kernel complexity changes for a $6 \times 6$ half-filled lattice. Comparison values are obtained with the Gutzwiller, Jastrow and RBM states, all using the same form for the orbital and symmetry-projected reference, while auxiliary-field QMC is able to provide unbiased benchmarks at half-filling \cite{Qin2016}.
The \red{bootstrapped GPS is `pretrained' via an initial RVM step on an optimized Jastrow wavefunction,} and demonstrates a systematic improvement with increasing data set size, surpassing the Gutzwiller accuracy with two data points.

As the number of data points increases, it can obtain a description of the same level as Jastrow with similar numbers of parameters, and can ultimately achieve a compact form with an energy error
\red{comparable to the RBM approach in a very systematic fashion.}
\red{Because we choose representations with real variational parameters, the accuracy is ultimately limited by the nodal surface constraints of the Pfaffian state}.
This could be further improved with standard backflow or Lanczos step technology \cite{PhysRevLett.122.226401, Pfau2019}, while in the future we will look to also describe the sign structure of these systems within the GPS model \red{which also represents a challenging problem for neural network quantum states} \cite{szabo2020neural, Choo2019, Bagrov19}.
However, even without these, the results demonstrate the convergence and improvability of the GPS relative to other approaches, and underline the advantage of this data-driven approach, where these correlated features can be modelled with very few parameters, for a sparse and accurate representation.

%
%
%
%
%
\begin{figure}
  \centering
  \includegraphics[width=\columnwidth]{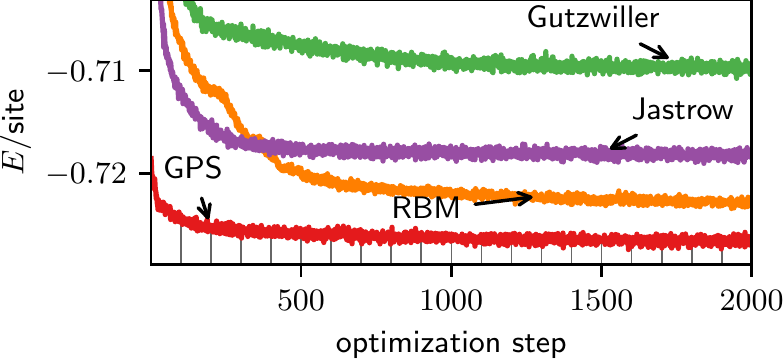}
  \caption{Convergence of the ground state energy for the $8 \times 8$ Hubbard model, at $U=8t$ and with 0.875 electrons per site.
  \red{The bootstrapped GPS representation is initialized at the optimized Jastrow wave function, resulting in rapid convergence.}
  Vertical grey lines indicate the iterations at which the configurational data set of the GPS is updated.
  \red{The convergence is shown for Gutzwiller (1 parameter), Jastrow (34 parameters), the RBM with $\alpha = 16$ (2064 parameters) and the bootstrapped GPS with optimized distance weights ($N_b + 63$ parameters).
  The number of data configurations specifying the GPS after each application of the RVM in the algorithm range from $\sim1000$ to $\sim2000$ configurations}.
  }
  \label{fig:bootstrapping_2D_8x8_doped_U_8}
\end{figure}

Finally, we consider the tricky $n=0.875$ hole-doped parameter regime of the model where agreement between the most accurate numerical methods is substantially harder to achieve compared to the half-filled case \cite{LeBlanc2015}.
Furthermore, no exact results are available for benchmarking, with many competing low-energy inhomogeneous phases claimed as the ground state \cite{Zheng2017, Zheng2016, PhysRevB.85.081110, PhysRevB.74.085104}.
We increase the lattice size to $8 \times 8$, and compare the convergence of the bootstrapped GPS to other variational approaches in Fig.~\ref{fig:bootstrapping_2D_8x8_doped_U_8}.
\red{We again initialize the GPS by a fit on the Jastrow wavefunction and use the $p \to \infty$ kernel with variationally optimized distance weights $f(d)/\theta$ to obtain the maximal flexibility of the representation.}

\red{While the RBM at $\alpha=16$ results in an appreciable improvement of the energy per site compared to the Jastrow ansatz, a further improvement by almost the same amount is achieved with the bootstrapped GPS which comprises a similar number of variational parameters as the RBM (although this number fluctuates as the data set selected changes).
Due in part to the efficient pre-training of the GPS on initial wave function data (in this case sampled from the optimised Jastrow ansatz), the GPS ansatz exhibits rapid and robust convergence. However, we are cautious of firm conclusions as to their ultimate relative accuracy and compactness, especially with the rapid developments of more advanced network architectures \cite{Liang2018, Levine2019,HibatAllah2020} and associated optimization techniques \cite{Roth2020} in the neural-network quantum state field.}


%
%
%
%

\section{Summary and Outlook}

We present a compact and efficient representation of quantum many-body wave functions, the Gaussian Process State. By allowing the parameters of the model to follow a statistical distribution, we derive a Bayesian inference framework for the probability amplitudes of the state in terms of an exponentially resummed space of physical correlation features. These are expressed in a kernel function which can compare any two configurations in this feature space, and allow for their amplitudes to be learnt from a data set, while also allowing insight into dominant features of this learnt state. The state can be deterministically and efficiently constructed with a manifestly compact data set from any initial state. As the data set is allowed to increase in size, the state is then shown to be able to universally approximate any entangled quantum state. \red{In this way, these states complement the neural-network states, also derived from machine learning principles to approximate any quantum state, and we anticipate many synergies between these approaches in the future. We also see the importance of the Bayesian framework as set out in this work in this endeavor, and the compact nature of the data-driven GPS approach as a key tool to reliably and systematically move beyond existing ansatze-driven approaches in strongly-correlated systems.}

We also present two numerical approaches which exploit this wave function form directly to approximate an unknown quantum state. In the first, the model is trained on amplitudes from a small subsystem, whose features are then implicitly extrapolated to define a state over the whole system. This is numerically simple, however the restriction to relatively short-ranged entanglement features inspired the development of the bootstrapped iterative optimization approach, which does not rely on any previous wave function information. This has broad applicability, and is shown to be efficient, sparse and accurate for both the half-filled and hole-doped strongly correlated regimes of the Hubbard model. We expect these schemes to work similarly well for other systems relevant for a broad range of scientific domains, and opens a new route for the compact representation of quantum many-body systems directly in terms of data.

\section*{Acknowledgements}
G.H.B. gratefully acknowledges support from the Royal Society via a University Research Fellowship, and funding from the Air Force Office of Scientific Research via grant number FA9550-18-1-0515.
A.G. acknowledges funding by the Engineering and Physical Sciences Research Council (EPSRC) through the Centre for Doctoral Training ``Cross Disciplinary Approaches to Non-Equilibrium Systems'' (CANES, Grant No. EP/L015854/1), and by the European Union through the MaX Centre of Excellence for supercomputing applications (Projects No. 676598).
The project has also received funding from the European Union's Horizon 2020 research and innovation programme under grant agreement No. 759063.
We are grateful to the UK Materials and Molecular Modelling Hub for computational resources, which is partially funded by EPSRC (EP/P020194/1).
A.G. thanks C. Zeni for useful discussions, while G.H.B. thanks Giuseppe Carleo, Stephen Clark and Garnet Chan for helpful discussions and comments on the manuscript.

A.G. and Y.R. contributed equally to this work.

\appendix
\section{Limits of the kernel function}
\label{subsec:SM_kernel_limits}

The kernel function is one of the key ingredients of the GPS ansatz proposed in this work since it implicitly models the entanglement and correlations that can be represented.
Its general form, repeated here for convenience,
%
\red{
\begin{align}
  k(\mathbf{x}, \mathbf{x}') &= \sum_{i}^L \delta_{x_i x_j'} \, \left( \frac{\tilde{k}(\mathbf{x}_i, \mathbf{x}'_j)}{\sqrt{\tilde{k}(\mathbf{x}_i, \mathbf{x}_i) \tilde{k}(\mathbf{x}'_j, \mathbf{x}'_j)}} \right)^p \label{eq:kernel_appendix}\\
	\tilde{k}(\mathbf{x}_i, \mathbf{x}'_j) &=  \theta + \frac{1}{p} \sum_{d \in D}  \delta_{x_{d(i)} x'_{d(j)}} f(d),
  \label{eq:kernel_local_appendix}
\end{align}
}
%
\red{depends on the two hyper-parameters:  $p$ and $\theta$ as well as the chosen set of displacements $D$ and the distance weighting function $f(d)$.}
The ``power'' $p$ controls the maximum rank of the correlations modelled (i.e., the maximum number of sites for which an explicit feature is present); the ``lengthscale'' $\theta$ allows tuning of the importance of accurately fitting low-rank compared to high-rank features in the ansatz; \red{the distance weighting function $f(d)$ controls the weighting of features depending on their range} and $D$ determines the set of displacements between two sites which are included in the correlations. In this work, this is generally chosen to be the set of all possible displacements.
It is instructive to look at the behaviour of the kernel for certain limiting cases of the hyperparameters just described, since these give rise to well known physical descriptions.
 %
%
\paragraph{Gutzwiller. Limit $\theta \rightarrow \infty$ or $p=0$.}
A Gutzwiller representation is recovered by either letting $\theta$ go to infinity or, equivalently, by taking $p$ to be zero.
Indeed, in both cases the kernel in (\ref{eq:kernel_appendix}) takes the very simple form
%
\red{
\begin{equation}
		k_G(\mathbf{x}, \mathbf{x}') = \sum_{i}^L \delta_{x_i x_j'}.
\end{equation}
}
%
\red{This kernel scans over the sites in the configuration $\mathbf{x}$ and counts the number of sites with occupancies identical to that of $\mathbf{x}'$ at reference site $j$. This reference site of the data in the training lattice can be chosen arbitrarily in translationally symmetric systems.}
It is straightforward to check that the above kernel can be written as \red{the scalar product in a feature space, which is then symmetrized over the translations of the system lattice, as  \mbox{$k_G(\mathbf{x}, \mathbf{x}') = \sum_i^L \boldsymbol{\Phi}^G_i(\mathbf{x}) \cdot \boldsymbol{\Phi}^G_j(\mathbf{x}')$}, where the feature vectors are defined as}
%
\red{
\begin{equation}
	(\boldsymbol{\Phi}^G_i)_s(\mathbf{x}) = \delta_{s,x_i} , 
\end{equation}
}
where $s$ is a specific occupancy $s \in \{ \cdot, \uparrow, \downarrow,\uparrow \downarrow \}$ and 
\red{$\boldsymbol{\Phi}^G_i(\mathbf{x})$ therefore simply checks whether site $i$ in ${\mathbf{x}}$ has a given occupancy $s$.}
\red{The choice to sum over the sites of the lattice ($i$) reflects the choice to symmetrize the features over the lattice, while the neglect of symmetrization over the training lattice ($j$) avoids additional cost in kernel evaluation, without substantial loss of accuracy.}
%
%
This means that the kernel $k^G$ implicitly defines a linear model on the features
\red{$\boldsymbol{\Phi}^G_i(\mathbf{x})$.}
This can be considered a generalised Gutzwiller form which favours or suppresses a given configuration depending on the number of sites in any given local occupancy, $s$.

\paragraph{Jastrow. Limit $\theta = 0$ and $p = 1$.}

A generalisation of the Jastrow representation is recovered by setting $\theta = 0$ and $p = 1$.
Indeed, for such a choice of hyperparameters the kernel takes the form
%
\red{
\begin{equation}
	k_J(\mathbf{x}, \mathbf{x}') = \sum_{i}^L \sum_{d \in D} \delta_{x_i x_j'} \,   \delta_{x_{d(i)} x'_{d(j)}} f( d ).
\end{equation}
}
As in the case of the Gutzwiller limit just discussed, it is also possible to write the above kernel as a \red{(symmetrized)} scalar product in some explicitly constructed feature space.
In this case the features \red{with respect to site $i$ and displacement vector $d_{12}$} are defined by the following rank-3 tensor
%
\red{
\begin{equation}
	(\boldsymbol{\Phi}^J_i)_{s_1,s_2,d_{12}}(\mathbf{x}) = \sum_l^L \delta_{s_1 x_i} \delta_{s_2 x_l} \delta_{d_{12} d_{il}} \sqrt{f( d_{il} )}.
\end{equation}
}
The component $s_1,s_2,d_{12}$ of
\red{$\boldsymbol{\Phi}^J_i$}
evaluates to \red{$\sqrt{f(d_{12})}$ whenever the occupancies of the site with index $i$ and the displaced site with index $d_{12}(i)$ are found with the local occupancies $s_1$ and $s_2$.}
Similarly to the the Jastrow factor this representation depends on two sites and their relative positions.
However,
\red{$\boldsymbol{\Phi}^J_i$} can be considered of generalized Jastrow form, since it depends on the specific occupation of the two sites, rather than their charge densities. It should also be stressed that the $\sqrt{f( d_{il} )}$ does not determine the overall amplitude of the feature, but rather the importance in trying to fit these to the data. Its overall weight is instead determined by the weights over the training configurations which exhibit that \red{correlation} feature.
%

\paragraph{Squared exponential kernel. Limit $p \rightarrow \infty $.}

Setting the hyperparameter $p$ to finite values is a physically meaningful way to restrict the space of correlations defined by the kernel function to a certain rank of correlations.
However, in certain cases (e.g., in the bootstrapping calculations described in this work) it might be convenient to take the $p \to \infty$ limit, obtaining a very flexible kernel controlled by the single hyperparameter $\theta$ \red{as well as the chosen distance weighting function $f(d)$}.
To evaluate this limit, we exploit the identity \mbox{$e^{x} = \lim_{n\rightarrow \infty}(1+\frac{x}{n})^n $}, obtaining the following kernel
%
\red{
\begin{eqnarray}
\label{eq:kernel_p_infinity}
		 k_{SE}(\mathbf{x}, \mathbf{x}') &=& \sum_{i}^L \delta_{x_i x_j'} \, e^{-\gamma_S^2(\mathbf{x}_i, \mathbf{x}'_j)/2 \theta} \\
	 \label{eq:kernel_p_infinity_distance}
	 \gamma_S^2(\mathbf{x}_i, \mathbf{x}'_j) &=& S(\mathbf{x}_i, \mathbf{x}_i) +S(\mathbf{x}_j', \mathbf{x}'_j) - 2S(\mathbf{x}_i, \mathbf{x}'_j)\\
	S(\mathbf{x}_i, \mathbf{x}'_j) &=& \sum_{d \in D}  \delta_{x_{d(i)} x_{d(j)}'} f( d ).
\end{eqnarray}
}
The above kernel can be seen as a \emph{squared exponential} kernel on the distance induced by the scalar product $S(\mathbf{x}_i, \mathbf{x}'_j)$.
Such a distance metric is related to the concept of a \emph{Hamming distance} \cite{Hamming1950} in discrete strings.
This squared exponential kernel includes by construction all possible correlation features, depending on a single hyperparameter $\theta$ which models the relative importance of fitting higher order to lower order features in the training data.
%

\paragraph{Entangled Plaquette States. Limit $p \rightarrow \infty $ and $\theta \rightarrow 0$.}
The Entangled Plaquette States ansatz (also known in the literature as correlator product states or complete graph tensor networks) decomposes the overall amplitude of a configuration into a product of contributions coming from a set of plaquettes over the degrees of freedom \cite{Changlani2009,Mezzacapo_2009}. Each possible state local to each plaquette is given its own variational parameter, with the number of parameters therefore growing exponentially with plaquette size.
Spanning the space of these features is recovered from the kernel in (\ref{eq:kernel_appendix}) in the limit of $p \rightarrow \infty $ and $\theta \rightarrow 0$, with the range of the effective plaquettes then being controlled by the allowed displacements in $D$.
We have shown that the limit $p \rightarrow \infty $ evaluates to Eq.~(\ref{eq:kernel_p_infinity}).
For $\theta \rightarrow 0$, the exponential $e^{-\gamma_S^2(\mathbf{x}_i, \mathbf{x}'_j)/2 \theta}$ in Eq.~(\ref{eq:kernel_p_infinity}) tends towards a delta function $ \delta^D_{\mathbf{x}_i \mathbf{x}_j'} $ evaluating to zero unless the local occupations of $\mathbf{x}$ around $i$ and $\mathbf{x}'$ around $j$ are identically equal, up to the set of displacements defined by $D$.
We can hence write the resulting kernel as
%
\red{
\begin{equation}
	k_{C}(\mathbf{x}, \mathbf{x}') = \sum_{i}^L \delta_{x_i x_j'} \delta^D_{\mathbf{x}_i \mathbf{x}_j'}.
\end{equation}
}
The combined Kronecker symbols $\delta_{x_i x_j'} \delta^D_{\mathbf{x}_i \mathbf{x}_j'}$ only evaluate to one if all the sites within a given cutoff in the two configurations are exactly the same and we can hence write it as $\delta_{(x_i, \mathbf{x}_i), (x_j',\mathbf{x}_j')}$.
Similar to the Gutzwiller and Jastrow limits, the above kernel can also be rewritten as an explicit scalar product \red{symmetrized across the lattice}.
In this case, the corresponding features are defined by the tensor
%
\red{
\begin{equation}
	(\boldsymbol{\Phi}^C_i)_{\mathbf{s}}(\mathbf{x}) = \delta_{\mathbf{s}, (x_i,\mathbf{x}_i)},
\end{equation}
}
where now the index $\mathbf{s}=(s_1,s_2,\dots)$ runs over all Fock states within the $|D|+1$ site plaquette around site $i$.
%
%
Linear regression on
\red{$\sum_{i=1}^L \boldsymbol{\Phi}^C_i(\mathbf{x})$} gives rise to an entangled plaquette states ansatz, favouring or suppressing a given amplitude depending on each \q{plaquette} of $|D|+1$ sites that is found within the corresponding configuration.
It is important to note here that while the parameterization of a \emph{direct} EPS ansatz grows exponentially with $|D|$ (roughly as $4^{|D|}$), the number of parameters when modelled \emph{indirectly} through the above kernel only scales with the size of the data set. The amplitudes of these features are then directly inferred from the weights over the test configurations.

\section{Representability of specific states}
\label{subsec:Representatbility}

{\bf W state:}
The GPS can also be used to represent certain many-qubit entangled states central to quantum information theory.
As an example, here we show how to represent the entangled W state of $n$ qubits,
\begin{equation}
  \ket{W} = \frac{1}{\sqrt{n}} (\ket{10 \ldots 0} + \ket{01 \ldots 0} + \ldots + \ket{00 \ldots 1}),
\end{equation}
with the GPS.
The W state can be represented using a single training configuration, namely one of the $n$ basis configurations which generate the state, e.g.
\begin{equation}
  \mathbf{x}'  = \ket{10 \ldots 0},
\end{equation}
and a kernel function with $p \to \infty$ and $\theta \to 0$ where the delta function only distinguishes between the two local occupancies $\ket{0}$ and $\ket{1}$.
With this choice, the kernel $k(\mathbf{x}', \mathbf{x})$ always either evaluates to $1$, if $\mathbf{x}$ is one of the $n$ translations of the configurations $\mathbf{x}'$ which create the W state, or to $0$ otherwise.
Setting the weight associated with the data configuration to the normalization $w_{\mathbf{x}'} = \ln \frac{1}{\sqrt{n}}$, we therefore obtain a GPS representation of the W state using only a single data configuration.

{\bf Laughlin wavefunction:}
Modifying the kernel function used also allows representation of other known ansatzes in very compact forms.
As an example, we consider the Laughlin wavefunction discretized on a real-space lattice, an ansatz describing $n$ spinless electrons on a two-dimensional lattice of $L$ sites.
Expressing the lattice position of site $k$ as a complex number $Z_k = x_k + i y_k$, where $x_k$($y_k$) is the $x$($y$) coordinate of the site, this discretized Laughlin wavefunction can be written as \cite{Changlani2009,Glasser2016}
%
%
\begin{equation}
  \Psi_L(\mathbf{x}) \propto \prod_{i < j }^n (Z_i - Z_j)^q e^{-\alpha \sum_{k=1}^n |Z_k|^2} ,
\end{equation}
where $q$ and $\alpha$ are fixed by the external model and desired state, with indices $i$, $j$ and $k$ referring to occupied sites in a sampled configuration ${\bf x}$.

We can express the Laughlin wavefunction as a GPS using a single data configuration with a modified kernel, or equivalently as a product of two GPS forms. This may be of interest as a starting point to build more flexible ansatzes for this problem. We define the modified kernel as
\begin{equation}
  k_{L}(\mathbf{x}, \mathbf{x}') = -\alpha \, k_{1B}(\mathbf{x}, \mathbf{x}') + k_{2B}(\mathbf{x}, \mathbf{x}'),
\end{equation}
which is a linear combination of the one and two-body kernels $k_{1B}$ and $k_{2B}$ (with $k_{1B}$ weighted by the parameter $\alpha$).
The one-body kernel is given as
\begin{equation}
  k_{1B}(\mathbf{x}, \mathbf{x}') = \sum_{k=1}^L  \delta_{x_{k} x'_{k}} \vert Z_k \vert^2,
\end{equation}
where the sum runs over all sites of the lattice.
This definition is equivalent to the local kernel $\tilde{k}$ of Eq.~(\ref{eq:kernel_local_appendix}) with $p=1$ and $\theta=0$, and with $f(d)=\vert d \vert^2$. 
To define the two-body kernel, we express the displacements in the local kernel as complex values $(d_x + i d_y)$, 
and use the kernel of Eq.~\ref{eq:kernel_appendix} with $p=1$, $\theta=0$, and a displacement weighting function of $f(d) = \ln(d_x + i d_y)^q$,
\red{where the site with index $0$ is selected as reference site in $\mathbf{x}'$}.
This gives a two-body kernel as
%
%
\begin{equation}
  k_{2B}(\mathbf{x}, \mathbf{x}') = \sum_{i}^L \delta_{x_{i} x'_{0}} \sum_{d \in D}  \delta_{x_{d(i)} x'_{d(0)}} \ln (d_x + i d_y)^q.
\end{equation}
Here the sum is taken over all possible displacements for which the displacement of site $i$ gives a site with larger index, i.e. $D = \{d : d(i) > i\}$.
Using a single data configuration, $\mathbf{x}'$, in which all sites are occupied $\mathbf{x}' = |1, 1, \dots, 1 \rangle$, the factor $e^{k_{2B}(\mathbf{x}, \mathbf{x}')}$, reproduces the two-body factor of the Laughlin wavefunction, $\prod_{i < j}^n (Z_i - Z_j)^q = e^{\sum_{i < j}^n \ln((Z_i - Z_j)^q)}$, and the factor $e^{-\alpha \, k_{1B}(\mathbf{x}, \mathbf{x}')}$ gives the one-body factor $e^{-\alpha \sum_{k=1}^n |Z_k|^2}$.
With the introduced definition of the kernel function $k_L$, the Laughlin wavefunction is therefore obtained as the GPS representation
\begin{equation}
  \Psi_L(\mathbf{x}) \propto e^{k_{L}(\mathbf{x}, \mathbf{x}')}.
\end{equation}

\section{Relevance Vector Machine}
\label{subsec:RVM}

In this appendix, we describe the Bayesian algorithms used to compute the GPS model parameters to optimally fit a target state. This includes the weights of the data points, the selection of the sparse configurational data set itself, as well as optionally the optimization of kernel hyperparameters. An expanded discussion of this GPS model optimization can be found in Ref.~\onlinecite{rath2020bayesian}.

The wave function ansatz in Eq.~(\ref{eq:GP_wavefunction}) can be rewritten as
\begin{equation}
	\ln \Psi_g(\mathbf{x}) = \sum_{b} w_b k(\mathbf{x},\mathbf{x}_b) \, ,
	\label{eq:GP_wavefunction_log}
\end{equation}
where the summation runs over all configurations in the data set $\{\mathbf{x}_b\}_{b=1}^{N_b}$.
The goal of the Relevance Vector Machine (RVM) is to find a sparse set of weights $\mathbf{w}$ from the training data which optimally represent the target state, as given by samples of its log wave function amplitudes.
\red{
As a standard modelling assumption, we model the \emph{likelihood} of the logarithm of the amplitudes as a Gaussian distribution
\begin{equation}
	p(\ln \boldsymbol{\Psi} \mid\mathbf{w}, \mathbf{\Lambda}) = \mathcal{N}(\mathbf{K}\mathbf{w} , \mathbf{\Lambda}),
	\label{eq:likelihood}
\end{equation}
where we have defined the matrix $\mathbf{K}$ of dimensions $N \times N_b$ and entries $K_{ib} = k(\mathbf{x}_i, \mathbf{x}_b)$.}

\red{Using a constant variance $\mathbf{\Lambda} = \mathbf{I}\sigma^2$ would allow to achieve a constant precision $\sigma^2$ on the logarithm of the amplitudes in Eq.~\eqref{eq:GP_wavefunction_log}. This does not fully encompass the aim we are trying to achieve, because a constant precision on the \emph{logarithm} of the wave function amplitudes does not correspond to a constant precision on the amplitudes themselves.}
As an example, a wave function with amplitudes ranging from $10^{-5}$ to \mbox{$10^{-1}$} will have amplitude logarithms (in base 10) ranging from $-5$ to $-1$.
Fitting all amplitudes to an accuracy of $\sigma^2 = 0.1$ means that, roughly speaking, the largest and the smallest amplitudes will be allowed to fluctuate between $10^{-1.1}$ and $10^{-0.9}$ and between $10^{-5.1}$ and $10^{-4.9}$ respectively.
Clearly these fluctuations are much larger for the larger amplitudes, meaning that they will be represented to a lower precision in the true amplitudes compared to the smaller amplitudes. This also results in the selected data set being less sparse than it could be, as small weighted amplitudes are comparatively being very finely resolved.

To overcome this problem, we can let each log amplitude, $\ln \Psi_i$, be represented to a different precision $\sigma_i^2$ and define the variance $\mathbf{\Lambda}$ in Eq. \ref{eq:likelihood} as $\Lambda_{ij} = \delta_{ij}  \, \sigma_i^2$.
The variances $\sigma_i^2$ can be chosen to precisely compensate for the fact that we are fitting to the logarithm of the state, by ensuring that the likelihood distribution of a given log wave function amplitude has a smaller variance if the magnitude of its expectation is larger.
The variance of a given amplitude, $\Psi_i$, under the assumption of Eq.~\eqref{eq:likelihood} that its logarithm is normally distributed, is
\begin{equation}
	\text{Var}(\Psi_i) = \langle \Psi_i \rangle^2 (e^{\sigma_i^2} -1)
\end{equation}
This demonstrates that the variance of $\Psi_i$ increases for increasing $\langle \Psi_i \rangle$, which we aim to compensate.
By assuming that the expectation value for the amplitude is the same as the one given by the data set, we can fix the variance of each amplitude as $\text{Var}(\Psi_i) = \tilde{\sigma}^2$, and solve for the amplitude-dependent precision in the likelihood, as
\begin{equation}
	\sigma_i^2 = \ln \left( 1+\frac{\tilde{\sigma}^2}{\Psi^2_i} \right).
\end{equation}
This allows us to define a constant variance, $\tilde{\sigma}^2$, to model all wave function amplitudes, and correspondingly use the configuration-dependent $\sigma_i^2$ above to model the variance in the likelihood distribution of specific values of $\ln \Psi_i$.

To enforce sparsity, RVM further models the \emph{prior} distribution of each parameter $w_b$ with a normal distribution centered around zero and an independent variance of $\alpha^{-1}_b$.
This gives rise to the following \emph{sparse prior} on the weights
\begin{eqnarray}
	p(\mathbf{w}\mid\boldsymbol{\alpha}) &=& \mathcal{N}(\boldsymbol{0},\mathbf{A}^{-1}) \label{eq:WeightPrior} \\
					\mathbf{A}	&=& \textrm{diag}(\boldsymbol{\alpha}).
\end{eqnarray}
Having specified prior and likelihood distributions, the \emph{posterior} distribution of the weights can be formally written down using Bayes' theorem as
\begin{equation}
	p(\mathbf{w} \mid \ln \boldsymbol{\Psi} , \mathbf{\Lambda}, \boldsymbol{\alpha} ) = \frac{p(\ln \boldsymbol{\Psi} \mid\mathbf{w},\mathbf{\Lambda}) p(\mathbf{w}\mid\boldsymbol{\alpha}) }{p( \ln \boldsymbol{\Psi}  \mid\boldsymbol{\alpha},\mathbf{\Lambda})}.
	\label{eq:Bayes_theorem}
\end{equation}
Since prior and likelihood are Gaussian, the posterior is also Gaussian and can be computed in closed form \cite{Tipping2000}
\begin{equation}
	p(\mathbf{w} \mid \ln \boldsymbol{\Psi} , \mathbf{\Lambda}, \boldsymbol{\alpha} ) = \mathcal{N}(\boldsymbol{\mu}, \mathbf{\Sigma}) ,
\end{equation}
with
\red{
\begin{eqnarray}
	\boldsymbol{\mu} &=& \mathbf{\Sigma} \mathbf{K}^{\rm{T}} \mathbf{\Lambda}^{-1} \ln \boldsymbol{\Psi} \\
	\mathbf{\Sigma} &=& ( \mathbf{K}^{\rm{T}} \mathbf{\Lambda}^{-1} \mathbf{K} + \mathbf{A})^{-1}
\end{eqnarray}}
The mean $\boldsymbol{\mu}$ of the posterior distribution is the best estimate for the value of the parameters $\mathbf{w}$. 
The value of the variances $\boldsymbol{\alpha}$ are optimized by a procedure called \emph{type-II maximum likelihood}.
This consists of the maximization of the denominator of Eq.~(\ref{eq:Bayes_theorem}), the \emph{marginal likelihood}, with respect to these variances on the prior of the weights and likelihood of reproducing the target state. The value of the marginal likelihood (or its logarithm) can also be given in closed form \cite{Tipping2003}. An efficient algorithm is used, whereby we initialize with an empty data set, and data is `greedily' added by ensuring the update to a parameter
$\alpha_b$ gives rise to the maximum increase in the log marginal likelihood. This also guarantees convergence to a local maxima within the RVM.
%
%
%
%
During this optimization, the sparse prior dictates that many $\alpha_b$ will be optimized at values of infinity, indicating that the basis configuration $\mathbf{x}_b'$ is not relevant to describe the wave function within the feature space of the kernel. This configuration can be ignored within the ansatz in Eq.~\eqref{eq:GP_wavefunction_log}, as its features are spanned effectively by the other data points.
%
%

%
We typically set $\tilde{\sigma}^2$ to a specific value to reach a desired accuracy and sparsity with the fit.
\red{Alternatively, we can determine a sensible choice for $\tilde{\sigma}^2$ for a given data set and kernel choice by running the RVM multiple times and maximizing the marginal likelihood with respect to the parameter $\tilde{\sigma}^2$, e.g. by the hyper parameter optimization technique presented in \cite{pmlr-v28-bergstra13} implemented in the hyperopt python package \cite{Bergstra_2015}.}
%

\red{Finally, we note that we typically want the likelihood variance $\tilde{\sigma}^2$ to be defined relative to the overall wave function scale, which will depend on the normalization of the state we fit to.
However, the optimized weights favor solutions distributed around zero due to the zero-mean prior of Eq.~\ref{eq:WeightPrior}. We can therefore renormalize the training data by shifting the log-amplitudes we fit on such that their mean vanishes.
If we are interested in describing the true amplitudes (including their normalization) with the learned GPS, we can simply add the rescaling constant of the training data back to the exponential for GPS predictions, effectively shifting the predicted log-amplitudes.
Because the shift of the log-amplitudes corresponds to a multiplication of the actual amplitudes, the normalization factor determined in the training can be ignored if the overall norm of the wave function is irrelevant, as is e.g. the case for Monte Carlo evaluation of the energy.
It was found that enforcing this zero-mean condition on the log-amplitudes of the training data helps to numerically stabilize the described bootstrapping procedure since the magnitude of the weights are well behaved, and are then commensurate with the imposed zero-mean prior on these quantities.}
%

\end{document}